\documentstyle[prl,aps,epsfig,preprint]{revtex}
\draft
\begin{document}	
\title{Self-assembled chains of graphitized carbon nanoparticles}
\author{A. Bezryadin, R. M. Westervelt, and M. Tinkham}
\address{Department of Physics and Division of Engineering and Applied
Sciences,\\Harvard University, Cambridge, Massachusetts 02138}
\date{December 21, 1998}
\maketitle
\begin{abstract}

We report a technique which allows self-assembly of conducting nanoparticles
into long continuous chains. Transport properties of such chains have been 
studied at low temperatures. At low bias voltages, the charges are pinned 
and the chain resistance is exponentially high. Above a certain threshold 
($V_{T}$), the system enters a conducting state. The threshold voltage is 
much bigger than the Coulomb gap voltage for a single particle and decreases 
linearly with increasing temperature. A sharp threshold was observed up to 
about $77 \ K$. Such chains may be used as switchable links in Coulomb charge 
memories.
\end{abstract}

\vspace{0.5cm}
%\begin{multicols}{2}

\pagebreak 

One-dimensional (1D) arrays of small metallic islands weakly coupled by 
tunneling are expected to have remarkable transport properties~\cite{Devoret}.
Uniform arrays can be used in single electron memories~\cite{Likharev} 
or in electron pumps which may allow a new metrological standard of 
capacitance~\cite{Keller}. Collective charge pinning is predicted for 
disordered arrays~\cite{MW}.
Such systems enter a conducting state only above a certain threshold voltage,
which increases with the number of islands in the array. Therefore 1D arrays 
can serve as switchable electronic links for Coulomb charge traps. High 
operational temperatures and good reproducibility could be achieved if the 
arrays are composed of nanometer scale metallic particles synthesized 
chemically (see for example Ref.~\cite{Schmid}). New approaches should 
be developed in order to organize such nanoparticles into useful
electronic devices.

This Letter describes a self-assembly technique which is used to arrange 
conducting nanoparticles into long continuous chains. The process of 
electrostatic self-assembly takes place between a pair of voltage biased 
microelectrodes, immersed in a dielectric liquid with suspended nanoparticles.
The electric field generated between the electrodes polarizes conducting 
particles and, due to the dipole-dipole attraction, leads to formation of a 
continuous chain which links the electrodes. Note that this chain formation 
effect~\cite{Halsey} is 
responsible for different electrorheological phenomena and was studied 
in macroscopic systems in this context~\cite{Havelka}. Here we demonstrate 
that this mechanism, if used on the micrometer scale, can produce continuous 
and electrically conducting chains of nanoparticles with very interesting 
and possibly useful transport properties. Our samples exhibit a Coulomb 
threshold behavior up to rather high temperatures ($T \sim 77 \ K$), even 
though we use relatively big particles of diameter $D \approx 30 \ nm$. 
Many of the properties of such self-assembled chains can be understood in the 
framework of the model of collective charge transport in
disordered arrays, proposed by Middleton and Wingreen (MW)~\cite{MW}.

A pair of leads bridged by a $\approx 1.2 \ \mu m$ chain of nanoparticles is 
shown in Fig.1a. The Cr electrodes (10 nm thick) have been fabricated using 
standard optical lithography and lift-off techniques. Longer chains are also 
possible as is shown in Fig.1b. In all cases we use ``onion'' type graphitized 
carbon nanoparticles~\cite{Carbon1}, of diameter $D \approx 30 \ nm$, 
available commercially~\cite{Carbon2}. We have verified in independent 
experiments that 
these particles are metallic down to at least $4.2 \ K$. For the self-assembly, 
a small amount of particles is dispersed ultrasonically in a dielectric 
liquid (toluene) until a uniform and slightly gray colored suspension is 
formed. An oxidized Si substrate with Cr electrodes is immersed into the 
suspension and the leads are connected to a $40 \ V$ voltage source in series 
with a big resistor $R_{s} = 1 \ G\Omega$. This resistor limits the current 
and serves to interrupt the process of self-assembly as soon as the 
first conducting chain is formed, similar to Ref.~\cite{Bezryadin}. 
Initially the current ($I$) between 
the electrodes is very low, $I<10 \ pA$. After a few seconds it increases 
suddenly by a few orders of magnitude. This indicates that the first continuous 
chain has been formed. Immediately after this current jump, the substrate 
is rinsed gently, dried and cooled down. Low noise current measurements have 
been carried out with an Ithaco 1211 current preamplifier. At low 
temperatures the noise was lower than $5 \ fA$ after averaging 
for $\approx 100$ s.

Current-voltage (I-V) characteristics (averaged over $\approx10$ scans) 
measured on one representative sample are plotted in Fig.2 for 
three different temperatures. The I-V curves are not hysteretic. 
At room temperature and low bias the I-V curve is linear with a 
zero-bias resistance of $\approx 12 \ G\Omega$. Already at relatively 
high temperatures ($\approx 77 \ K$) we observe a fully developed gap.
Within the accuracy of our measurements ($\approx 5 fA$) the current
is zero below a certain threshold voltage ($V_{T}$) which increases
with decreasing temperature. In the examples of Fig.2 the threshold is 
$V_{T} \approx 0.18 \ V$ at $T = 72 \ K$ and it increases up to 
$V_{T} \approx 0.38 \ V$ when the sample is cooled down to  
$T = 13 \ K$. I-V curves measured on six $L \approx 1.2 \ \mu m$ 
samples were similar to those shown in Fig.2. The absolute 
value of the threshold is subject to strong sample-to-sample 
fluctuations, probably due to irregularities in the chains. 
The averaged value of the threshold was found to be 
$\langle V_{T} \rangle = 0.3 \pm 0.2 \ V$ at $T = 4.2 \ K$. 
Measurements on a longer chain suggest that $V_{T}$ scales linearly 
with the chain length, as expected from the MW model. For a chain of 
$L \approx 5.6 \ \mu m$ we have found 
$V_{T} \approx 1.1 \ V$ (at $T=4.2 \ K$) while the expected value
(for a 4.7 times longer chain) is $\approx 1.4 \pm 0.4 \ V$. 
(The rms fluctuations $\Delta V_{T}$ are assumed to satisfy the 
usual relation $\Delta V_{T} \sim L^{1/2}$ \cite{MW}.)

The large value of the measured threshold voltage is a {\it collective}
effect. Indeed, if it would be due to a single very small particle 
in the chain (which would have a very big Coulomb gap), then the 
threshold should be observed even at $T=300 \ K$ because the thermal
energy at $300 \ K$ is only $k_{B}T \approx 26 \ meV$, what is much 
lower than $eV_{T} \approx 300 \ meV$. Experimental I-V curves do 
not show any gap at $300 \ K$ (Fig.2). 

It is instructive to compare the threshold voltage with the 
MW model which predicts that $I=0$ for $V<V_{T}$ where $V_{T}$
is proportional to the number of particles N. If the mutual 
capacitance between the particles $C$ is much smaller than 
the self-capacitance (or capacitance to the gate) $C_{0}$ then 
$V_{T} \approx Ne/2C_{0}$. 
In our case $C_{0} \approx 2 \pi \varepsilon_{0} D \approx 
1.7 \cdot 10^{-18} \ F$. To estimate $C$ we assume that the 
idealized polyhedron shaped particles of graphitized carbon~\cite{Carbon1} are 
assembled in the chain in such a way that the facets of neighbor particles are
parallel to each other. Therefore $C \approx A \varepsilon_{0}/d$ where 
$A=3 \sqrt{3}/2 (D/4)^{2} \approx 1.5 \cdot 10^{-16} \ m^{2}$
is the area of each facet. The distance $d$ between the facets can be 
approximated by the spacing between graphene layers in graphite,
so $d \approx 3.35$ \AA \ and finally $C \approx 4 \cdot 10^{-18} \ F$. 
(Therefore the charging energy of a single particle is 
$e^{2}/2(2C+C_{0}) \approx 8 meV$.)
Since $C$ is of the same order of magnitude as $C_{0}$, we have to use 
numerical results of MW:  $V_{T}C_{0}/eN \approx 0.09$ for $C/C_{0} = 2.35$.
With $N \approx L/D \approx 40$ as estimated geometrically, 
this gives $V_{T} \approx 0.34 \ V$ which is in good agreement with 
the experimental value $V_{T} = 0.3 \pm 0.2 \ V$ averaged
over six samples.

Another MW prediction is that above the threshold, when the voltage 
is high enough to populate the entire chain with electrons, the current 
follows a power law $I \sim ((V-V_{T})/V_{T})^{\zeta}$ where $\zeta=1$ 
and $5/3$ in the 1D and 2D cases, respectively (MW numerical simulation 
for the 2D case gives a different value of $\zeta \approx 2$). 
Experimentally we do observe a power-law behavior (see Fig.3a). 
The smaple-dependent exponent is $\zeta \geq 1$. Note that exponents 
bigger than unity ($\zeta \approx 1.4$) were measured previously on artificial 
1D arrays~\cite{Clarke}. 

This type of power-law I-V dependence suggests that a true transport threshold 
exists in our chains. This conclusion can be strengthened further if 
one can confirm experimentally that {\it below} the threshold the current 
is indeed suppressed. The idealized MW model predicts a strictly zero 
current for $V<V_{T}$. Such a prediction can not be verified experimentally. 
On the other hand, in any real system the current will not be exactly zero.
It should rather be suppressed exponentially below the threshold. 
This is true insofar as the dynamic state is separated from the static state 
by an energy barrier which can be overcome due to thermal or quantum 
fluctuations. (A process which leads to a nonzero subgap current 
is illustrated in the schematic of Fig.2.) An exponential I-V dependence 
should occur below the threshold if such a barrier appears at $V=V_{T}$ and then 
increases with $(V_{T}-V)$ when the voltage is decreased. We indeed observe
such exponential tails in our samples. Fig.3b illustrates that the at very low
currents (typically lower than $1 \ pA$) the I-V dependence is exponential,
while above the threshold it exhibits the power-law behavior as it was discussed
above.

Another important property of our samples is a relatively slow, 
namely linear (see Fig.4), dependence of the threshold on 
temperature~\cite{Deviation} .
The temperature dependence of $V_{T}(T)$ can be 
qualitatively understood in the following way.
At $T=0$, a non-zero current appears at $V=V_{T}$ when 
the voltage on a single junction is $V_{1}(0)=V_{T}(0)/N$
on average. This is just enough for electrons to overcome
the Coulomb barriers between the particles. 
At $T>0$, thermal fluctuations help overcome the 
barrier, so current appears at a lower voltage 
$V_{1}(T) \approx V_{1}(0)-k_{B}T/e$. Therefore the threshold voltage 
for the {\it chain} is reduced to 
$V_{T}(T) \approx V_{T}(0) - Nk_{B}T/e$, and $dV_{T}/dT = -Nk_{B}/e$
should be compared with the experimental value $-3.6 \ mV/K$ (Fig.4).
Inserting the estimate $N \approx 40$, one obtains 
$-3.5 \ mV/K$, in good agreement with experiment.
 
In conclusion, we have developed a new technique which is used to 
self-assemble nanoparticles into long continuous chains.
The electrons in the chains are collectively pinned at low voltages
by Coulomb energies. At the threshold voltage (which decreases 
linearly with temperature) the chains switch into a conducting state. 
The Coulomb blockade was observed up to $T \approx 77 \ K$.

This work was supported in part by NSF Grants DMR-94-00396, DMR-97-01487,
and PHY-98-71810.

\begin{figure}[h]
\vspace{0.1cm}
\end{figure}
\parbox[t]{14cm}{\small FIG.1 \ \ (a) \ An SEM micrograph of a chain of 
graphitized carbon nanoparticles self-assembled between two Cr microelectrodes.
The chain length is $L \approx1.2 \ \mu m$. (b) \ An example of a longer chain. 
The image shows a part of a $\approx 6 \ \mu m$ chain and one Pt 
electrode.}

\begin{figure}[h]
\vspace{0.1cm}
\end{figure}
\parbox[t]{14cm}
{\small FIG.2 \ \ Low current parts of three I-V curves measured on a 
$L \approx 1.2 \ \mu m$ chain at $T = 300, 72$ and $13 \ K$. The $300 \ K$ 
curve is divided by a factor of 20. The continuous curves at $V>0$ are the 
same exponential fits as in Fig.3b. The schematic on top shows a 1D array of 
islands (open squares) biased with a voltage which is slightly below the 
threshold. The number of induced extra electrons on each island is shown 
schematically by black dots above each island. The charge distribution 
is linear~\cite{MW}. This is similar to the flux distribution in type-II 
superconductors with strong pinning which are described by the Bean model.}

\begin{figure}[h]
\vspace{0.1cm}
\end{figure}
\parbox[t]{14cm}
{\small FIG.3 \ \ (a) \ Log-log plots of the current vs normalized 
voltage for three different samples of $L \approx 1.2 \ \mu m$ at 
$T=4.2K$. The bottom curve is shifted down by one decade for clarity. 
The linear fits are $I=I_{0}((V-V_{T})/V_{T})^{\zeta}$ 
where $\zeta =$ 1.03, 2.06, and 2.32 and respectively 
$I_{0}=0.15 \ nA$, $0.35 \ nA$, and $5.9 \ nA$. 
Corresponding threshold voltages are $V_{T} \approx$ $0.3 \ V$, 
$0.615 \ V$, and $0.385 \ V$.
(b) \ Semi-log plots of I-V curves measured 
at $T = 72, 45$ and $13 \ K$ on the same sample. Straight lines represent 
the exponential fits $I = \frac{V}{R_{0}}exp(V/V_{0})$. The parameters 
are $V_{0}= 36, 40$, and $48 \ mV$, $R_{0} = 5.1\cdot 10^{15}, 7\cdot 10^{16}$, 
and $1.9\cdot 10^{17}  \Omega$ at $T = 72, 45$ and $13 \ K$, respectively.}

\begin{figure}[h]
\vspace{0.1cm}
\end{figure}
\parbox[t]{14cm}{\small FIG.4 \ \ Temperature dependence of the 
threshold voltage measured on the same sample as Fig.2. The linear 
fit is $V_{T}(T)=V_{T}(0) - T\cdot dV_{T}/dT$ where $V_{T}(0)=0.41 V$ 
and $dV_{T}/dT=3.6 mV/K$.}   

\pagebreak 

\begin{figure}[h]
\vspace{3cm}
\centerline{ \epsfxsize=8cm \epsfbox{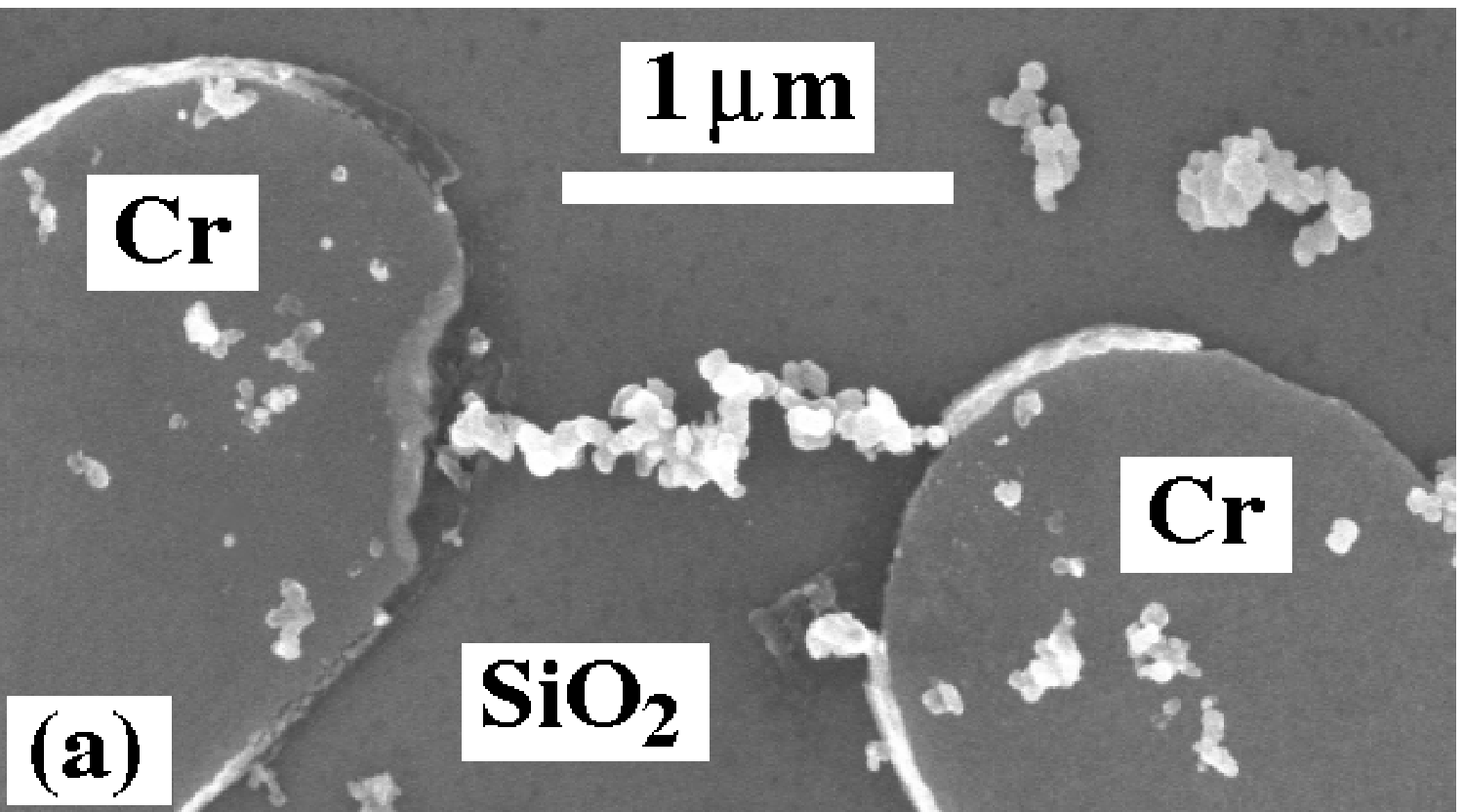}}
\vspace{0.1cm}
\centerline{ \epsfxsize=8cm \epsfbox{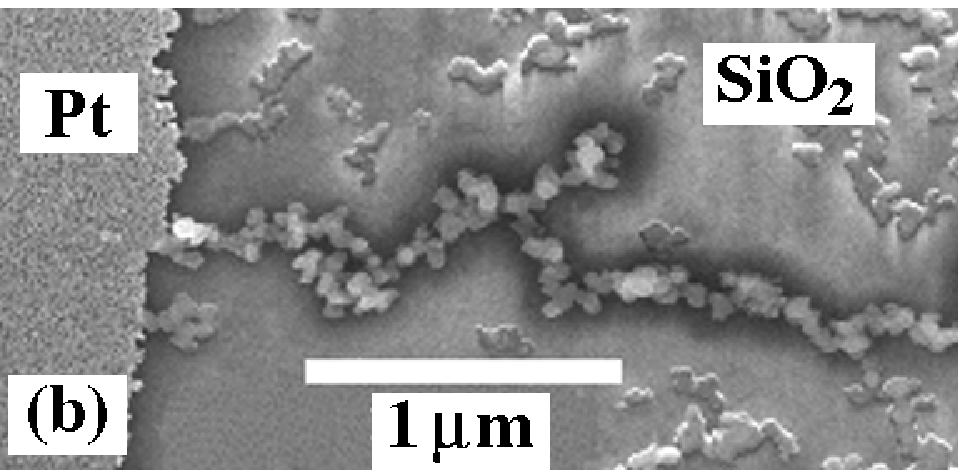}}
\end{figure}
\vspace{7cm}
\parbox[t]{14cm}{\small FIG.1 \ \ A. Bezryadin et al., Applied Physics 
Letters} 

\pagebreak 

\begin{figure}[h]
\centerline{ \epsfxsize=8cm \epsfbox{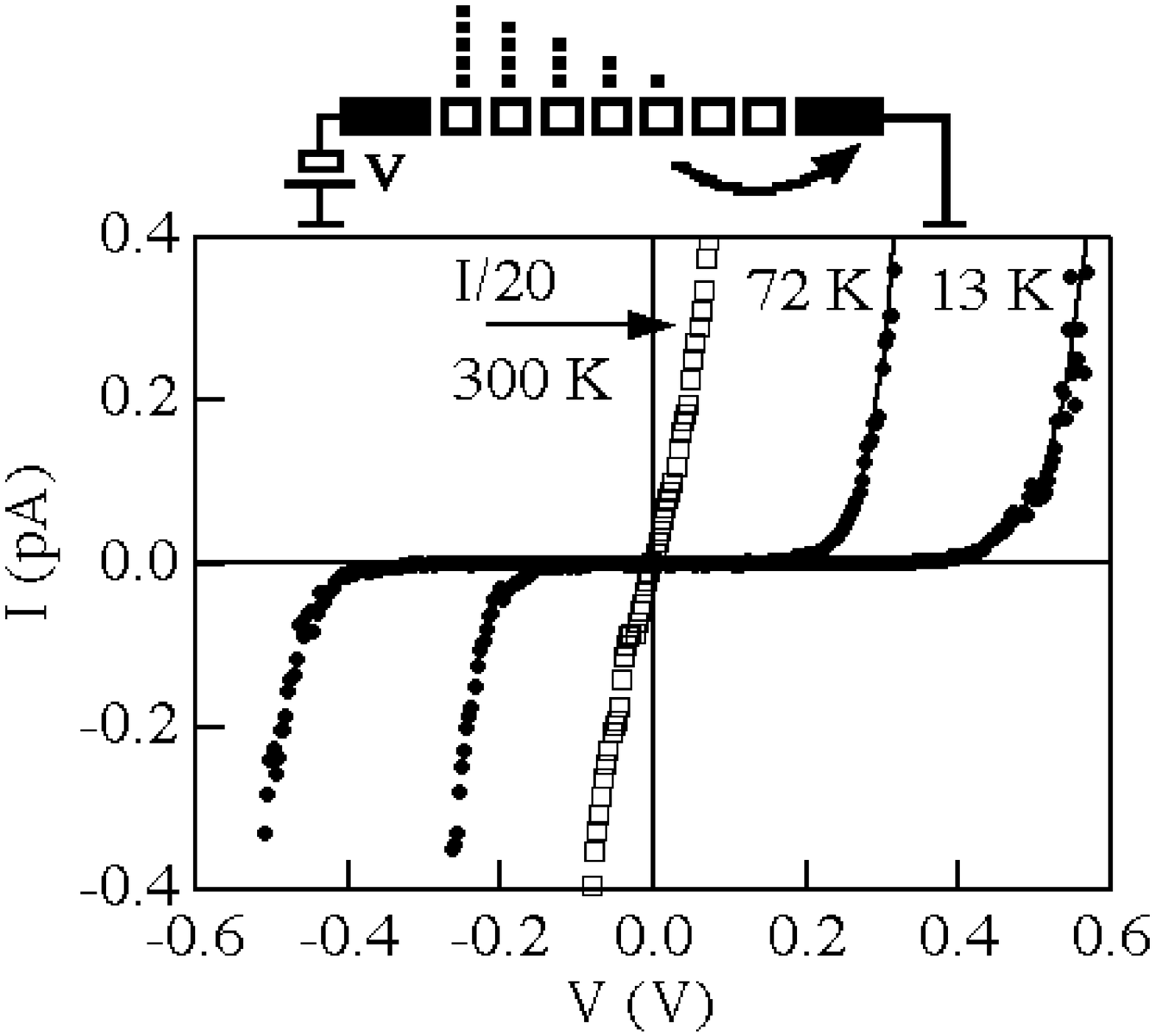}}
\end{figure}
\vspace{7cm}
\parbox[t]{14cm}
{\small FIG.2 \ \ A. Bezryadin et al., Applied Physics 
Letters}
\vspace{0.5cm}

\pagebreak 

\begin{figure}[h]
\centerline{ \epsfxsize=9cm \epsfbox{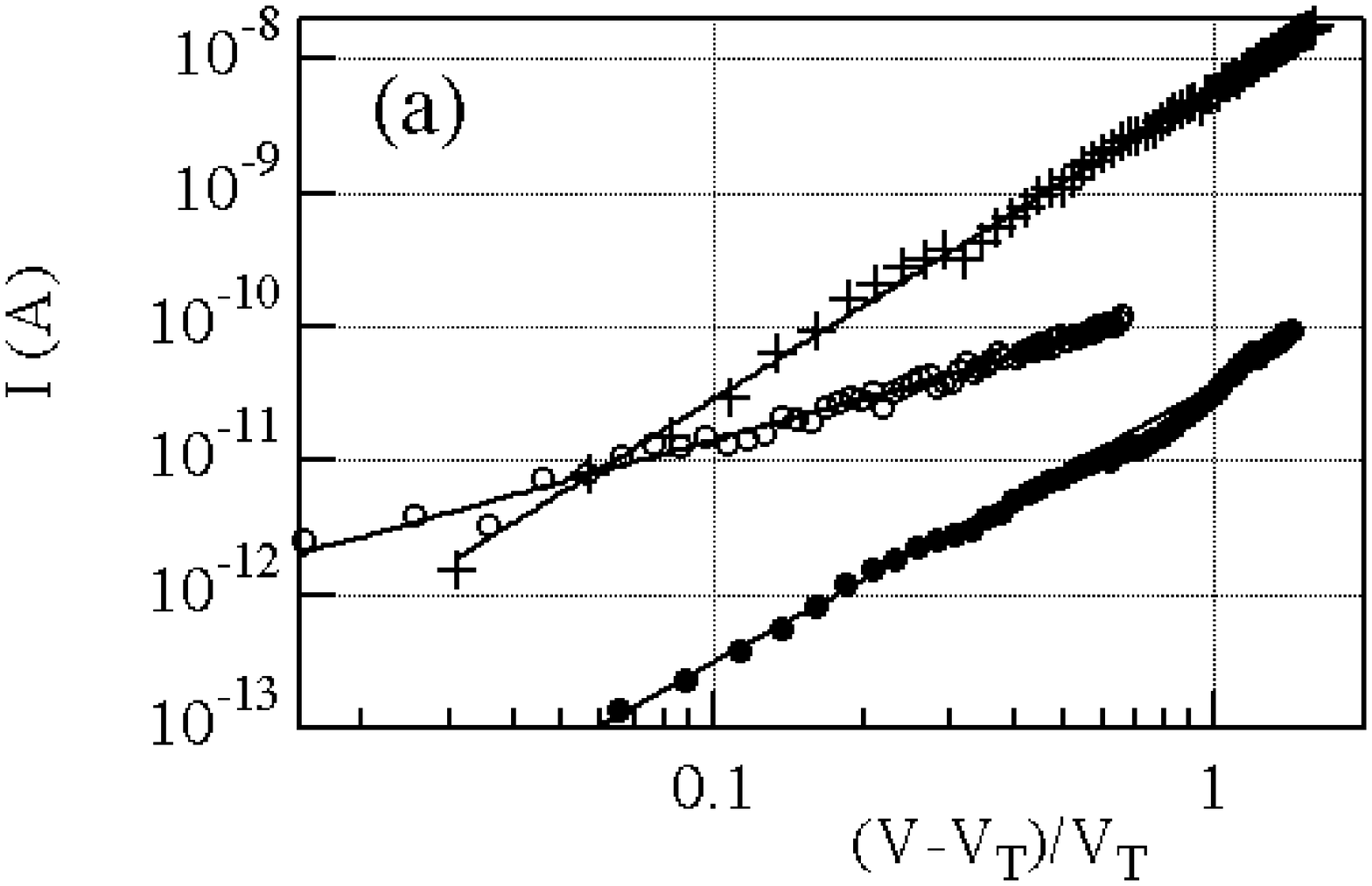}}
\vspace{0.1cm}
\centerline{ \epsfxsize=9cm \epsfbox{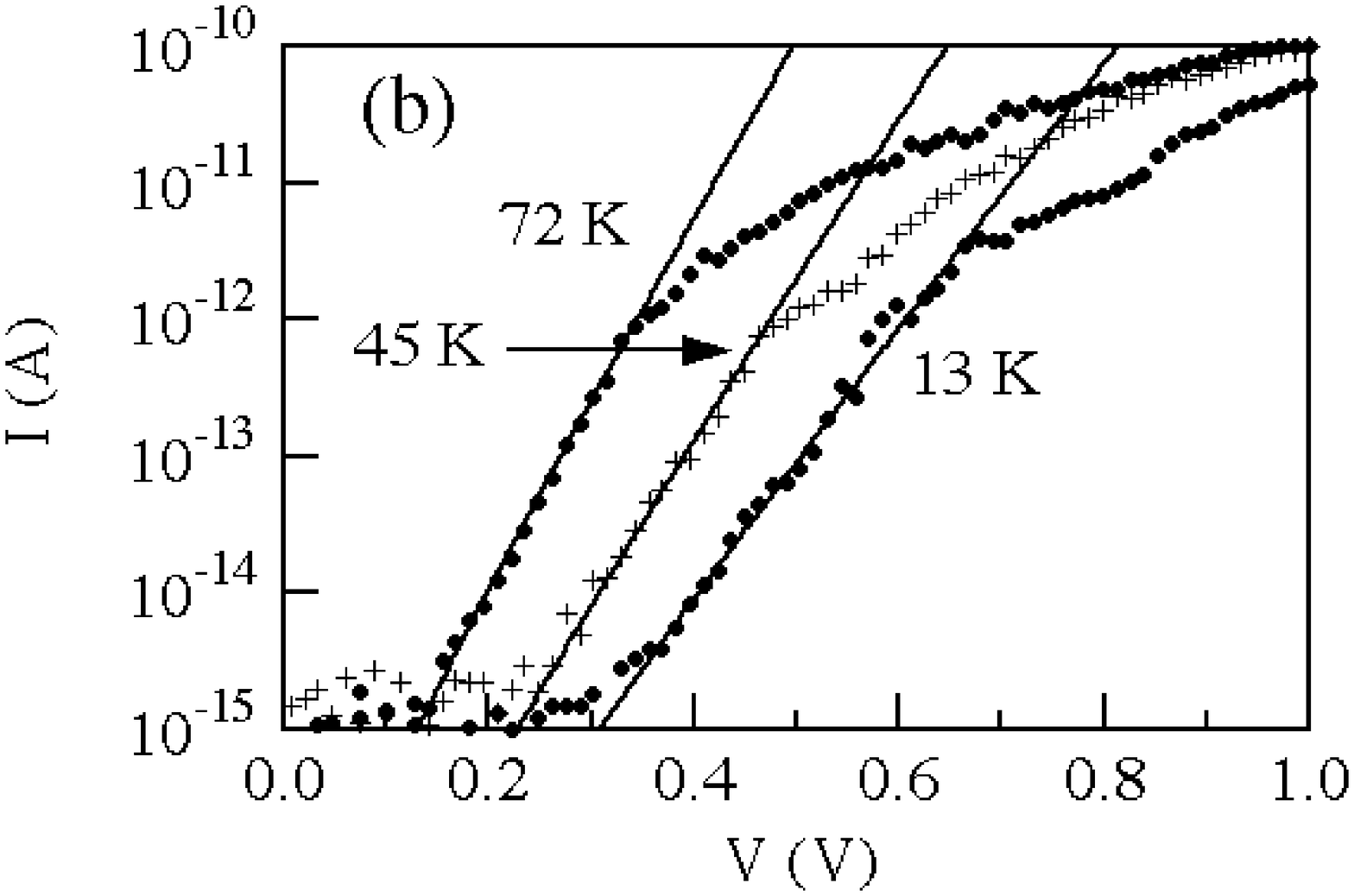}}
\end{figure}
\vspace{7cm}
\parbox[t]{14cm}
{\small FIG.3 \ \ A. Bezryadin et al., Applied Physics 
Letters}  

\pagebreak 

\begin{figure}[h]
\centerline{ \epsfxsize=8cm \epsfbox{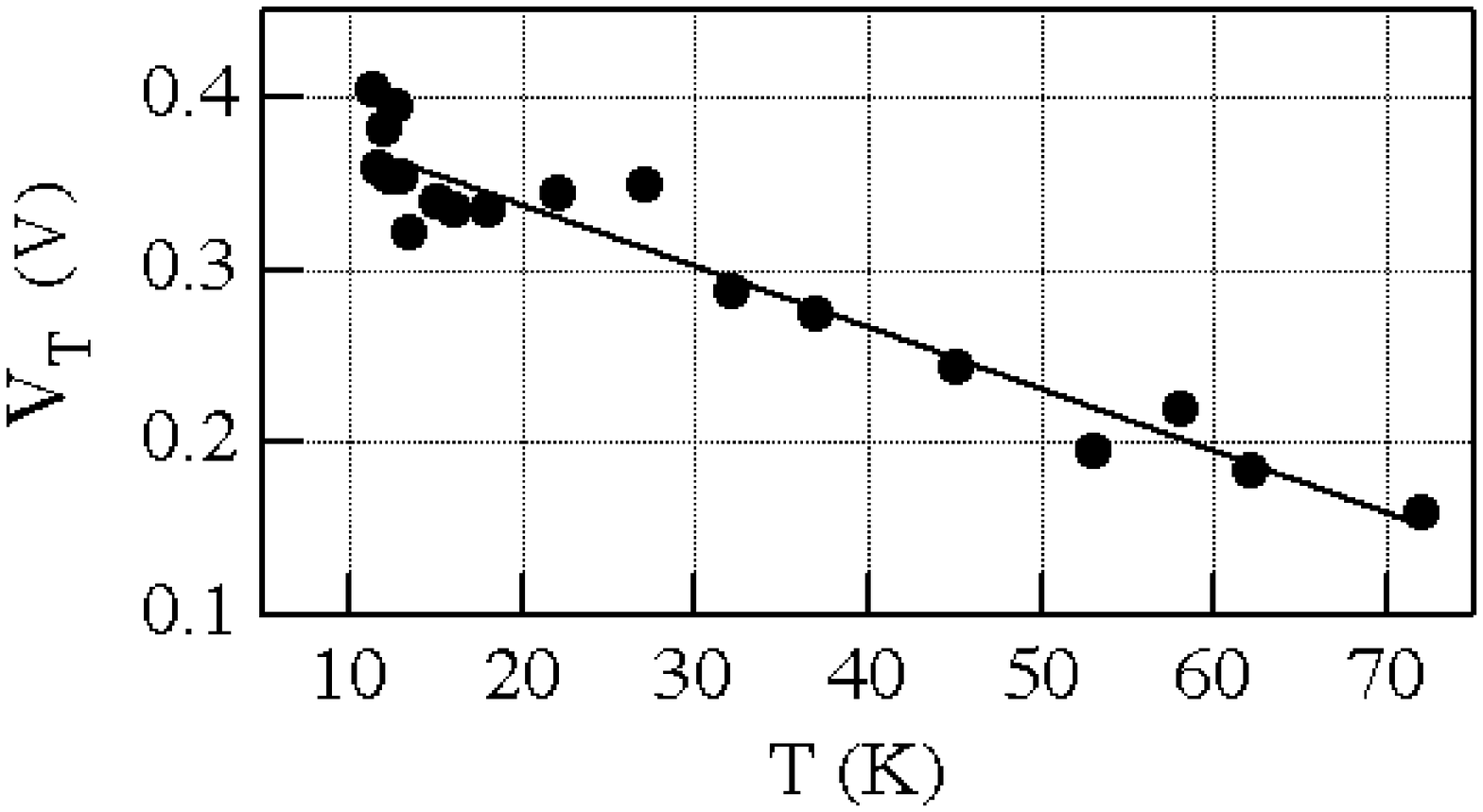}}
\end{figure}
\vspace{7cm}
\parbox[t]{14cm}{\small FIG.4 \ \ A. Bezryadin et al., Applied Physics 
Letters}   
\vspace{0.3cm}

%\end{multicols}

\begin{references}
\bibitem{Deviation}Each data point in Fig.4 is determined from an 
averaged I-V curve measured at a certain temperature. The absolute 
value of the threshold voltage fluctuates somewhat from one voltage 
scan to another, even if the temperature does not change. 
We attribute the deviations from the linear behavior, seen in Fig.4,
to these fluctuations.
\end{references}
\end{document}